\documentclass[11pt,a4paper]{article}
\usepackage{hyperref}
\usepackage[latin1]{inputenc}
\usepackage{dsfont}
\usepackage{graphicx}

\begin{document}
\title{Manipulation of Spherical Droplets on a Liquid Platform Using Thermal Gradients}
\author{Ehsan Yakhshi-Tafti, Hyoung J. Cho, Ranganathan Kumar \\
\\\vspace{6pt} Mechanical, Materials and Aerospace Engineering, \\ University of Central Florida, Orlando, FL 32765, USA}
\maketitle
Fluid Dynamics Video:\\
In the recent years, there has been a growing interest in droplet-based (digital) microfluidics for which, reliable means of droplet manipulation are required. In this study we demonstrate thermal actuation of droplets on liquid platforms, which is ideal for biochemical microsystems and lab-on-chip applications because droplets can be transported with high speed, good control and minimal thermal loading as compared to using conventional solid substrates. In addition, other disadvantages of using solid surfaces such as evaporation, contamination, pinning, hysteresis and irreversibility of droplet motion are avoided. 
\\
Based on the theoretical development and measurements, a silicon-based droplet transportation platform was developed with embedded Titanium micro heaters. A shallow liquid pool of inert liquid (FC-43) served as the carrier liquid. Heaters were interfaced with control electronics and driven through a computer graphical user interface. By creating appropriate spatio-temporal thermal gradient maps, transport of droplets on predetermined pathways was successfully demonstrated with high level of robustness, speed and reliability.
\\
The video shows normal imaging of droplet manipulation accompanied by the corresponding infrared thermal imaging showing the spatio-temporal temperature maps and the outline of the drop as it moves towards hot spots.
\\
\\

E. Yakhshi-Tafti, H. J. Cho, and R. Kumar, Applied Physics Letters 96, 264101 (2010).
\end{document}